\documentclass{mn2e}
\usepackage{url} 
\usepackage{amsmath}
\usepackage{amssymb}
\usepackage{graphicx}
\usepackage{epsfig}
\usepackage{ulem}
\usepackage{color}
\usepackage{IEEEtrantools}

\newcommand\lsim{\mathrel{\rlap{\lower4pt\hbox{\hskip1pt$\sim$}}
        \raise1pt\hbox{$<$}}}
\newcommand\gsim{\mathrel{\rlap{\lower4pt\hbox{\hskip1pt$\sim$}}
        \raise1pt\hbox{$>$}}}

\newcommand{\dd}{\partial}

\newcommand{\Msol}{{\rm M}_{\odot}}

\newcommand{\apj}{ApJ}

\newcommand{\aap}{A$\&$A}

    \def\dd{\partial}

    \def\beq{\begin{equation} }
    \def\eeq{\end{equation} }
    \def\spose#1{\hbox to 0pt{#1\hss}}
    \def\ltsim{\mathrel{\spose{\lower.5ex\hbox{$\mathchar"218$}}
     \raise.4ex\hbox{$\mathchar"13C$}}}

\def\spose#1{\hbox to 0pt{#1\hss}}
\def\lta{\mathrel{\spose{\lower 3pt\hbox{$\mathchar"218$}}
        \raise 2.0pt\hbox{$\mathchar"13C$}}}
\def\gta{\mathrel{\spose{\lower 3pt\hbox{$\mathchar"218$}}
        \raise 2.0pt\hbox{$\mathchar"13E$}}}

\begin{document}
\label{firstpage}

\title[Flares in GRBs]
{Flares in Gamma-Ray Bursts: Disc Fragmentation and Evolution}
\author[]{Simone Dall'Osso$^1$, Rosalba Perna$^1$, Takamitsu L. Tanaka$^1$,
Raffaella Margutti$^{2, 3}$\\
$^1$ Department of Physics and Astronomy, Stony Brook University, Stony Brook, NY, USA\\
$^2$ Center for Interdisciplinary Exploration and Research in Astrophysics (CIERA) and Department of Physics and Astrophysics \\
~~~Northwestern University, Evanston, IL 60208\\
$^3$ Center for Cosmology and Particle Physics, New York University, 4
Washington Place, New York, NY 10003, USA}

\maketitle
\label{firstpage}


\begin{abstract}

Flaring activity following gamma-ray bursts (GRBs), observed in both
long and short GRBs, signals a long-term activity of the central
engine. However, its production mechanism has remained elusive.  Here
we develop a quantitative model of the idea proposed by Perna et
al. of a disc whose outer regions fragment due to the onset of
gravitational instability.  The self-gravitating clumps migrate
through the disc and begin to evolve viscously when tidal and shearing
torques break them apart.  Our model consists of two ingredients:
theoretical bolometric flare lightcurves whose shape (width, skewness) is
largely insensitive to the model parameters, and a spectral correction to match 
the bandpass of the available observations, that is calibrated using the observed 
spectra of the flares. This simple model reproduces, with excellent agreement, the empirical
statistical properties of the flares as measured by their
width-to-arrival time ratio and skewness (ratio between decay and rise
time).  We present model fits to the observed lightcurves of two
well-monitored flares, of GRB~060418 and of GRB~060904B. To the best of our knowledge, 
this is the first quantitative model able to reproduce the 
flare lightcurves and explain their global statistical properties.
\end{abstract}

\begin{keywords}
gamma rays: bursts --- accretion, accretion discs --- X-rays: general --- 
black hole physics
\end{keywords}

\section{Introduction}

Over a decade of observations by the {\it Swift} satellite (Gehrels et
al. 2005) have opened a new window on the phenomenology of gamma-ray
bursts (GRBs).  Arguably one of the most unexpected phenomena has been
the discovery of erratic flaring activity in the X-ray luminosity of
the sources. These X-ray flares have been observed over a timescale of
seconds to even over a day after the prompt emission (e.g. Burrows et
al. 2005; O'Brien et al. 2005; Margutti et al. 2010), hence
overlapping with the timecale of the afterglow.

The standard model for the afterglow, that is the external shock
model, predicts a lightcurve characterized by a powerlaw or broken
powerlaws (e.g. Meszaros \& Rees 1997), depending on the regime of
observation. Several phenomena can cause departures from simple
powerlaws, such as inhomogeneities in the external medium (Lazzati et
al. 2002; Heyl \& Perna 2003), refreshed shocks due to the collision
of a late shell of plasma with the external shock material (Rees \&
Meszaros 1998), angular inhomogeneities in the fireball energy
distribution (Nakar et al. 2003) and gravitational lensing (Loeb \&
Perna 1998).  However, based on the observed properties of the flares,
Lazzati \& Perna (2007) demonstrated that---for at least the subset of
them with a ratio between duration to arrival time $\Delta t/t_{\rm
  p}\lesssim 0.25$--- inhomogeneities in  the external
shock cannot reproduce the observed properties of the flares. Rather,
these must be the result of long-term activity of the GRB central
engine.

Understanding the origin of the flares is of
great interest because it would provide insight into the inner workings
and physics of the GRB engines. The flares also provide an intriguing link
between long and short GRBs (e.g. Margutti et al. 2011a), which are attributed to different
progenitors, a collapsing star for the former (MacFadyen \& Woosley
1999), and a compact binary merger for the latter (e.g. Narayan et al. 1992). 

A number of ideas have been proposed over the years. For example, for the case of long GRBs, King et al (2005) 
suggested that the X-ray flares could be produced from the fragmentation of the collapsing stellar core 
in a modified hypernova scenario, while Lazzati et al. (2011) found that 
propagation instabilities of a jet piercing through a massive star might explain the properties
of the subset of flares with low luminosity contrast. For the case of short
GRBs, Dai et al. (2006) proposed a new mechanism of binary merger that
can incorporate the presence of a flare. However, the fact that the
flares are observed with similar properties in long and short GRBs (Margutti et al. 2011a)
makes particularly compelling the models based on what is generally
believed to be the common element in the two scenarios: an
accretion disc\footnote{Alternative ideas for a central engine have
  invoked a magnetar, which can be be produced either as a result of a
  supernova (Dall'Osso et al. 2011; Metzger et al 2011) or the merger
  of two neutron stars (Giacomazzo \& Perna 2013). However, even with a magnetar, 
a disk may still be present.}. Proga \& Zhang
(2006) and Giannios (2006) proposed mechanisms involving magnetic
instabilities, while Perna, Armitage \& Zhang (2006, PAZ in the
following) suggested that flares can be produced in an accretion disc
which, at large radii, fragments or otherwise suffers large-amplitude
variability. This idea was motivated by the early observations
(O'Brien et al. 2005; Cusumano et al. 2006) which indicated a positive
correlation between duration and arrival time of the flares, as
expected from clumps which form and viscously evolve from different
regions of the disc.

After several more years of {\it Swift}
observations, the properties of the flares have now been well
characterized, both from a statistical point of view, as well as
individually (see in particular Margutti et al. 2010). Some
similarities have also emerged with a subset of the early pulses
generally attributed to the prompt emission (Norris et al. 2005). 
The similarity in the distribution of waiting times of X-ray flares and 
prompt $\gamma$-ray pulses led Guidorzi et al. (2015)
to conclude that both phenomena represent the same underlying 
physical process, and to suggest disc fragmentation as the possible common origin.

Here, motivated by the wealth of recent observations which allow a
more quantitative assessment of theoretical models, we develop in
further detail the idea that a fragmented disc may give rise to the
observed flaring activity. After identifying the regions in the disc
that are subject to gravitational instability and subsequent fragmentation
(Sec.~2), we solve the time-dependent equation for the viscous
evolution of a clump, compute the bolometric lightcurve of a flare,
and study the dependence of its peak luminosity, duration and shape on the
physical parameters of the systems -- such as the mass of the clump and
its initial radius, the disc viscosity, and the mass of the central
object (Sec.~3). We generalize this study in Sec.~4, by investigating
how the varying physical conditions in the disc may affect the arrival
times of the flares. Then in Sec.~5 we perform a statistical
comparison of the properties predicted by our model with the
observations, as well as model in detail two bright flares which have
been well monitored. To the best of our knowledge, this is the first
time that the lightcurves of flares have been reproduced by means of
a specific theoretical model. We summarize and conclude in Sec.~6.

\begin{figure*}
\includegraphics[scale=0.50]{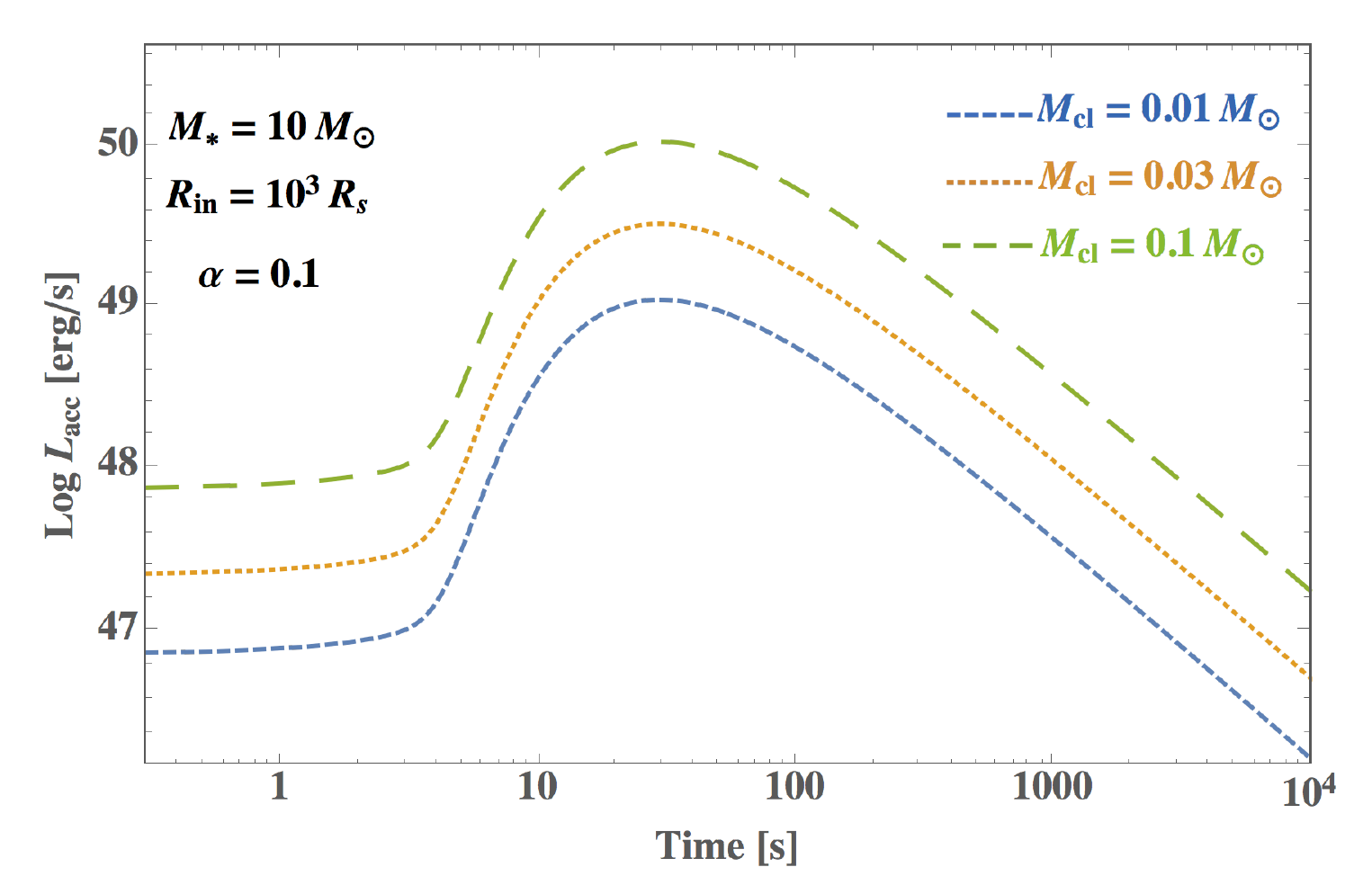}
\includegraphics[scale=0.50]{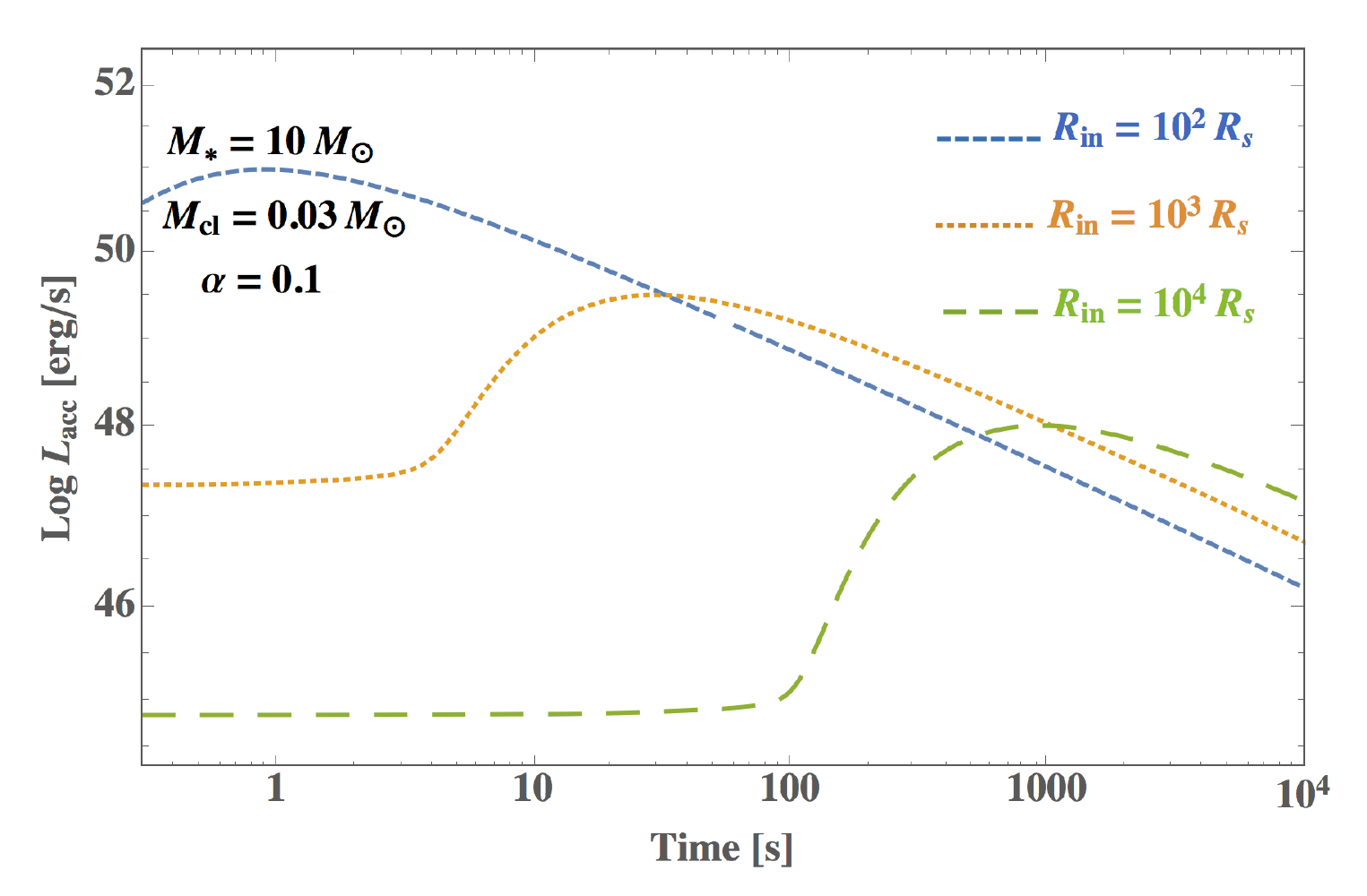}\\
\includegraphics[scale=0.50]{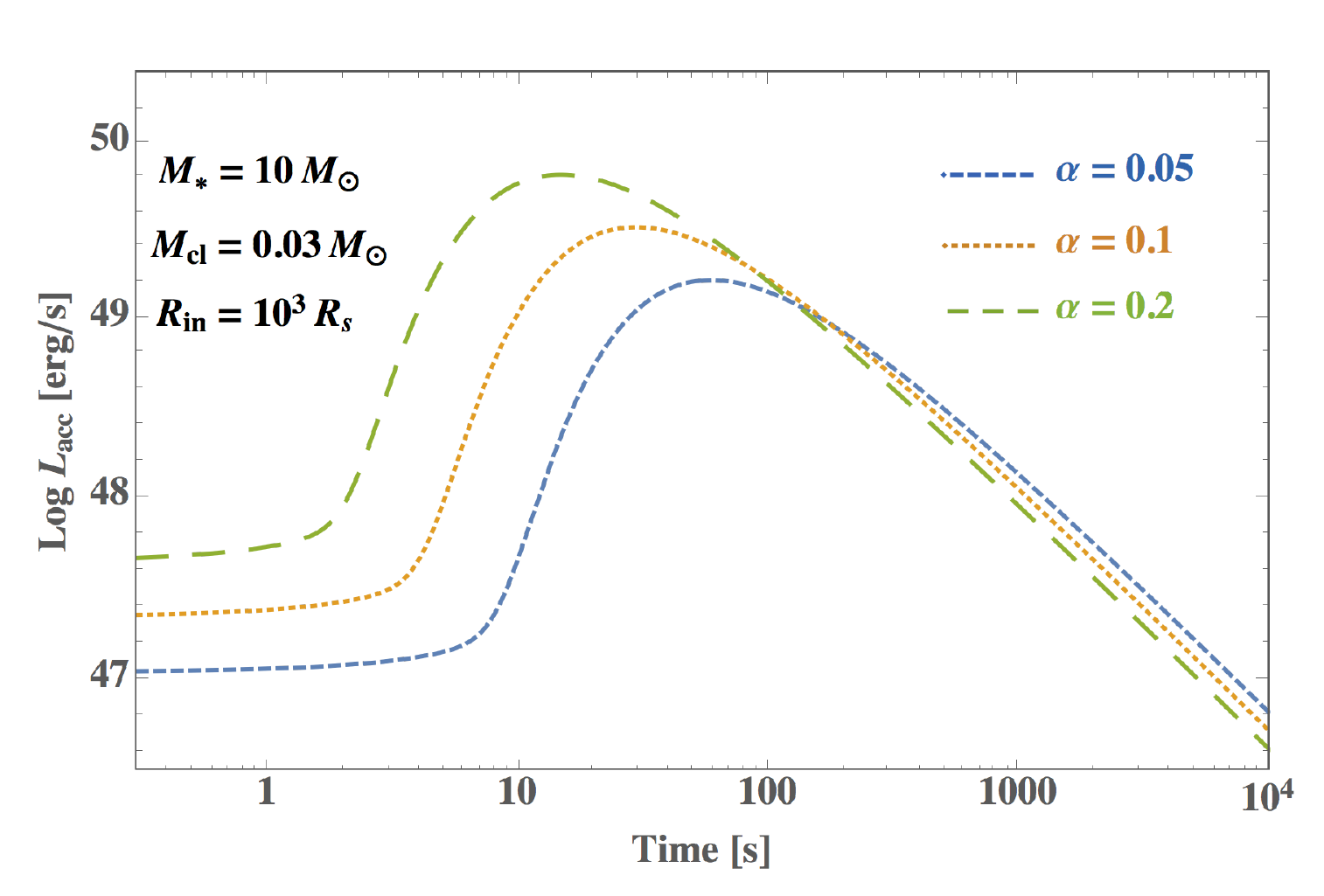}
\includegraphics[scale=0.50]{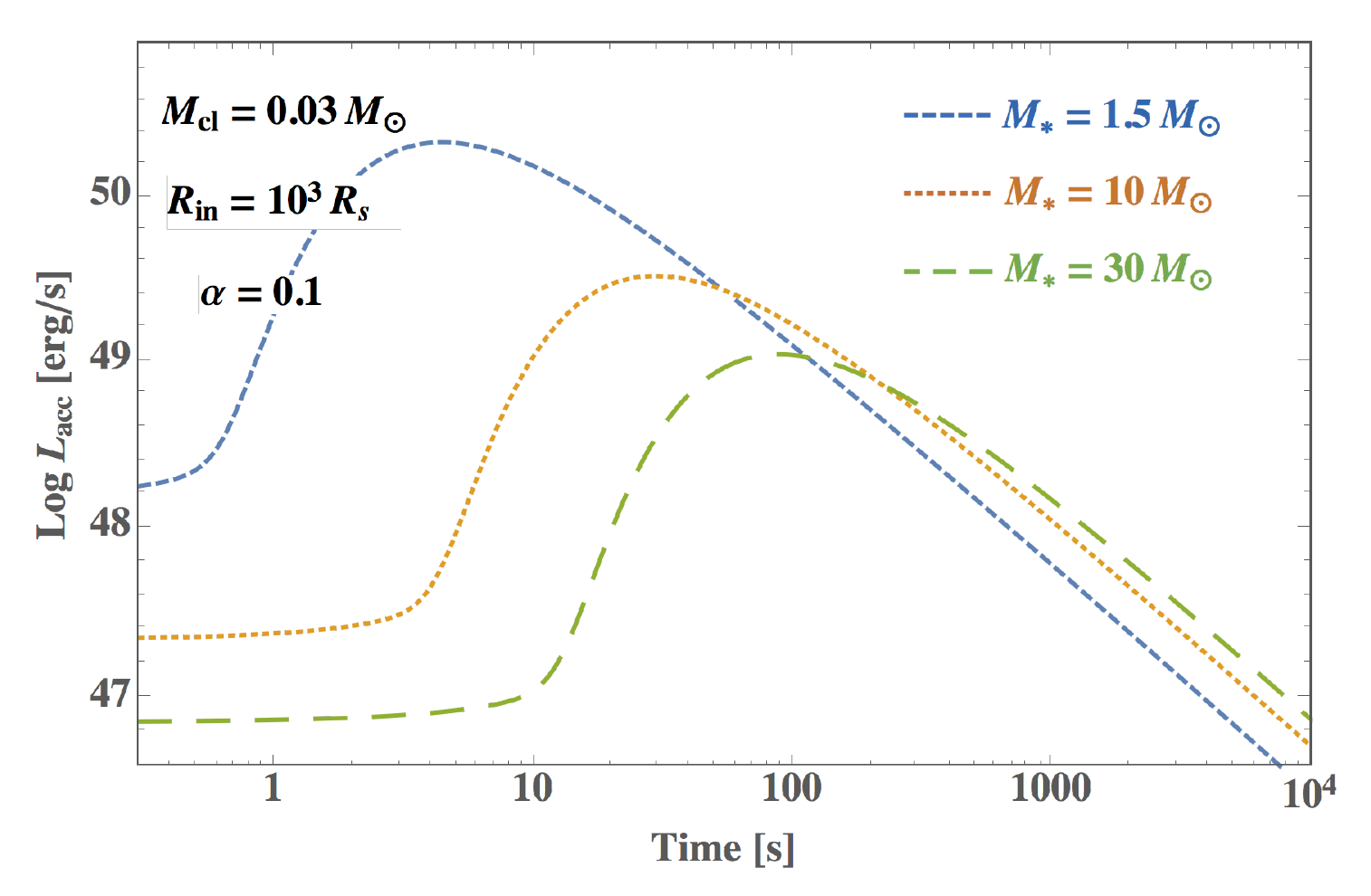}
\caption{Bolometric lightcurve of a flare for a combination of model parameters:
clump mass $M_{\rm cl}$ and initial location $R_{\it in}$ in the disc, magnitude of the viscosity
parameter $\alpha$, and mass $M_*$ of the central accreting compact object. Note that $t=0$ is defined as
the time of the GRB trigger which, in these examples, coincides with the time at which viscous evolution
of the clumps begins. }
\label{fig:case1}
\end{figure*}

\section{Fragmentation and Clump Formation in a disc}

As discussed in Sec~1, the idea by PAZ that GRB flares
may be due to viscously evolving blobs in a fragmented disc was partly
motivated by numerous findings that the outer regions of an
hyperaccreting disc are gravitationally unstable. The instability
sets in when the Toomre parameter (Toomre 1964)
\begin{equation}
Q_{\rm T} = { {c_{\rm s} \kappa} \over {\pi G \Sigma} } < 1,
\end{equation}
where $c_{\rm s}$ is the sound speed, $\kappa$ the epicyclic frequency,
$G$ is the gravitational constant, and $\Sigma$ is the disc surface density.
Di Matteo et al. (2002) and Chen \& Beloborodov (2007) found that, for accretion rates $\dot M \sim
10~\Msol $~s$^{-1}$, the disc becomes unstable at radii $R\gsim
50~R_{\rm s} \equiv R_{\rm inst}$, where $R_{\rm s}=2GM/c^2$ is the Schwarzschild
radius and $c$ is the speed of light.  The inner boundary of the unstable region, $R_{\rm inst}$,
increases with decreasing accretion rate, and becomes $\gsim 1000 R_{\rm s}$
for $\dot M \sim 0.1~\Msol$~s$^{-1}$. 
Recently, Liu et al. (2014) have extended the analysis of the gravitational instability to include
the dependence of the Toomre parameter on the vertical structure of the disk. They confirmed that the
instability sets in in the outer parts of the disk, and additionally they found that it is more easily produced 
further away from the midplane.

Once the disc becomes unstable, fragmentation into bound objects
will occur if  the local cooling time,
\begin{equation}
 t_{\rm cool} < t_{\rm crit} \approx 3 \Omega_{\rm K}^{-1}\;,
\end{equation}
where  $\Omega_{\rm K}$ is the local Keplerian angular velocity.
When the above condition is satisfied,
the maximum stress obtainable from gravitational
instability produces
insufficient heating to offset cooling (Gammie 2001; Rice et
al. 2003).  As discussed by PAZ, this condition is likely realized
under the extreme conditions of GRB discs. A further in-depth analysis
of cooling mechanisms in GRB discs was performed by Piro \& Pfhal
(2007).  They demonstrated that, at disc radii for which $Q_{\rm T}\sim 1$,
photodisintegration is a very effective coolant that promotes fragmentation.

Once a disc becomes unstable and fragmentation ensues, the mass of a
fragment is expected to be on the order of $\Sigma\lambda ^2$, where
$\lambda \sim Q_{\rm T}H$ is the wavelength of the fastest growing mode
(Binney \& Tremaine 1987), and $H$ is the disk scale height.  Hence, $M_{\rm frag}\sim \Sigma
(Q_{\rm T} H)^2$. However, fragments are expected to merge and/or accrete until
their tidal influence becomes strong enough to open a gap in the disc.
This happens when the mass of the fragment has grown to
\begin{equation}
 M_{\rm frag} \simeq \left( {H \over R} \right)^2 \alpha^{1/2} M_*\,, 
\end{equation}
where $M_*$ is the mass of the central object (Takeuchi et al. 1996), and
$\alpha$ the viscosity parameter in the disk (Shakura \& Sunyaev 1973).
Given that at the high accretion rates of GRB discs $H\sim R$, and
considering a range of masses between $M_*\sim 1.5-10~\Msol$ it is
reasonable to assume clumps with masses on the order of
$0.05-1~\Msol$.

In the following, under the assumption that the onset of the
gravitational instability in the outer regions leads to
self-gravitating clumps, we compute the time-evolution and the resulting
luminosity of the accreting matter.

\section{Viscous Evolution of self-gravitating Clumps}
\label{sec:3}

A clump can be modeled as a sharp concentration of mass at some radius
$R_{\rm in}$ (e.g. a delta function).  It will arrive at the center
(location of the compact object) and increase the luminosity at a
timescale 
\beq t_{\rm flare}\sim \beta t_{\nu}(R_{\rm in}) \sim \beta
\frac{R_{\rm in}^2}{\nu}, 
\eeq 
where $\nu$ is the viscosity coefficient, and
$\beta$ is a dimensionless factor whose exact value depends on the
initial profile and viscosity prescription (e.g. Milosavljevic \&
Phinney 2005, Tanaka \& Menou 2010).

We compute the viscous evolution of the clump semi-analytically
by solving the standard equation for axisymmetric accretion discs,
\beq
\frac{\dd}{\dd t}\Sigma (R,t)=\frac{1}{R}\frac{\dd}{\dd R}\left[ R^{1/2}\frac{\dd}{\dd R}\left(3\nu\Sigma R^{1/2}\right)\right].
\eeq
When the viscosity is a powerlaw of radius $\nu \propto R^n$,
the evolution of the surface density profile $\Sigma (t; R)$
can be solved exactly by  the integral
\beq
\Sigma(t; R) = \int_{R_{\rm i}}^{\infty} G(t; R, R^{\prime}) \Sigma(t=0; R^{\prime}) ~{\rm d}R^{\prime},
\label{eq:sigma}
\eeq 
where the Green's function $G$ depends on the boundary conditions\footnote{The main advantage of the Green's function method is that
it allows for the calculation of disk evolution and  accretion power with a single integration, for arbitrary initial surface density profiles.}
Solutions with inner boundary conditions
(e.g. the no-torque condition at the 
the innermost stable circular orbit
or a suppressed accretion rate due to  gravitational
torques or a magnetic field)
imposed at the cylindrical coordinate origin $R_{\rm i}=0$ were derived by Lust (1952) and Lynden-Bell
\& Pringle (1974); these solutions have divergent disc luminosities.
Solutions with $R_{\rm i}>0$, with $t\rightarrow \infty$ disc
luminosities matching those found in steady-state solutions, were
derived by Tanaka (2011).

Once $\Sigma(R;t)$  is known, the accretion power 
$L_{\rm acc}(t)$ due to the viscous spreading of the clump can be approximated as
\beq
L_{\rm acc}(t)\approx \int_{R_{\rm i}}^{\infty} \frac{9}{4}~\Sigma(R,t)~\nu(R)~\Omega_{\rm K}^2(R)~2\pi R~{\rm d}R\,.
\label{eq:Loft}
\eeq
Examples of viscosity prescriptions with the requisite form $\nu \propto R^n$
include isothermal discs ($\nu \propto R^{3/2}$, i.e. $n=3/2$) and
advection-dominated discs ($n\approx 1/2$; $c_{\rm s}\sim v_{\rm K}$;
$H\sim R$; e.g. Tanaka 2013).
Since the latter is more appropriate to the early accretion phases of interest here,
we will adopt this scaling law. 

The high-energy emission in GRBs is believed to be produced within a relativistic outflow (at the distances at which it 
becomes optically thin), and hence a crucial ingredient of GRB models is the ability to launch such jets. Two main mechanisms have been 
proposed: neutrinos (Popham, Woosley \& Fryer 1999, Ruffert \& Janka 1999) and magnetic fields (Blandford \& Znajek 1977, Blandford \& Payne 1982). 
As it is found that neutrinos might play an important role only at the highest mass accretion rates (e.g. Janiuk et al. 2007, Janiuk et al. 2013, 
Just et al. 2016), we will not consider this mechanism here. Instead, we will focus on scenarios where the rotational energy of a spinning BH is 
extracted by a large scale magnetic field threading the disk (Blandford \& Znajek 1977). This mechanism is commonly 
considered as the most viable 
for launching jets, and predicts a jet power that is directly proportional to
the mass accretion rate\footnote{Apart, possibly, from the very early
  phases associated to the prompt emission (Tchekhovskoy, Narayan \&
  McKinney 2010).} (Blandford \& Znajek 1977, Krolik \& Piran 2011,
2012; see also Kumar, Narayan \&  Johnson 2008). The jet power is then
converted into radiation farther along its path, e.g. through
collisions of shells with different velocities (Burrows et al. 2005;
Fan \& Wei 2005; Zhang et al. 2006, Kumar \& Zhang 2015). As long as
the radiative mechanism maintains a constant efficiency throughout the
duration of an individual flare, we can assume that the bolometric
luminosity of the flare will be $L_{\rm bol} = f_{\rm rad} L_{\rm
  acc}$, where $f_{\rm rad}$ - the efficiency of energy conversion
from mass accretion into radiation - is a constant for that event, but
may change among different events.

\section{Flares due to self-gravitating clumps: evolution under different physical conditions}
\label{sec:viscous-spreading}

Once a clump has been formed, it will eventually evolve viscously
according to Eq. (6). In this work, we envisage two main scenarios for
the time when viscous spreading begins.

\subsection{Early flares: prompt viscous spreading}
\label{sec:early}
The first scenario we consider is one in which clumps form during the
time that the prompt emission is produced, in regions of the disc
where viscous stresses are strong enough to shred the clumps right away.  In
this case, viscous evolution begins without significant delay with
respect to the GRB prompt.  The condition for this to happen will be
discussed in more detail in Sec.4.2.

Examples of solutions to Eq.~(\ref{eq:sigma}) are shown in Fig.~\ref{fig:case1} for
a range of values of the clump mass, the viscosity parameter, the disc radius at which 
the clump evolution starts ($R_{\rm in}$), and the mass of the 
accreting object.  To include the possibility that the central engine is a neutron 
star, we also considered a mass value of $1.5~\Msol$.
The figure shows the bolometric luminosity, computed via using Eqs.~(\ref{eq:sigma}) and (\ref{eq:Loft}).
We note that, as expected, the mass of the clump essentially determines the magnitude
of the flare, while the initial radii, the viscosity
parameter, and the mass of the central object influence
the peak time and the width.
These then affect the peak luminosity, since the total radiated energy $E \propto M_{\rm cl}$ is the
same for a given clump mass.
More specifically, we find 
that the peak luminosity of the flares scales with their initial position as 
(roughly) $R_{\rm in}^{-1.4}$, the peak time scales as (roughly)
$R_{\rm in}^{1.6}$, and the luminosity scales with the peak time as (roughly)
$t_{\rm p}^{-0.90}$.

These bolometric lightcurves are characterized by a ratio of the flare ``width''
to peak time on the order of $\Delta t/t_{\rm p}\approx 4.3$.  Here the width 
$\Delta t$ is defined as the interval between the {times  $t_1$ and $t_2$ 
(the former during the rise and the latter during the decay) at which the flux is 1/$e$ of the 
peak flux.
If, in addition, we define the rise time as the difference $t_{\rm r} = t_{\rm p} - t_1$ and the decay time as 
$t_{\rm d} = t_2 - t_{\rm p}$, then the bolometric lightcurves are also characterized by a 
ratio $t_{\rm d}/t_{\rm r}$, which measures the asymmetry (or skewness) of the curve.
We systematically find $t_{\rm d}/t_{\rm r} \approx 5.7$. 
These quantities are defined in the same way as the observational characterizations of
Margutti et al. (2010), to allow for direct comparisons (see Sec.~5). 

We note and emphasize an important property of our model lightcurves:
{\it the ratios $\Delta t/t_{\rm p}$ and $t_{\rm d}/t_{\rm r}$ are
  almost insensitive to the value of model parameters.}  The
robustness of the light curve shape at early times is due to the fact
that as long as the initial position of the clump is far away from the
inner boundary condition imposed at $R_{\rm ISCO}$, the early viscous
  evolution simply scales with the local viscous time profile
  ($\propto R^2/\nu(R)\propto R^{2-n}$) of the clump.  At late times,
  $L_{\rm bol}(t) \propto t^{-4/3}$ if $n=1/2$ (i.e. if $H/R\sim{\rm
    constant}$), with the decay powerlaw depending on $n$ (see Tanaka
  2011).  Increasing the initial distance of the clump increases the
  timescales of the early-time evolution by a factor $2^{2-n}$ but
  otherwise the surface density evolution is the same.  Similarly,
  $L_{\rm bol}(t)$ simply scales with the normalization of $\Sigma$,
  so changing the clump mass in our model does not change the shape of
  $L(t)$.\footnote{ Note that this is not true generally if the
    viscous properties of the disc scale non-linearly with
    $\Sigma$. In that case, changing the clump mass or the initial
    clump location could lead to different light curve shapes.}  We
  have also verified that replacing the $\delta$-function initial
  profile with Gaussian or step-function annuli does not strongly
  affect the shape of the curve.  Hence, the shape of the bolometric
  light curves can be considered as distinctive, predictive signature
  of the model.

These properties} will be discussed in
more detail and compared to the data in Sec.~5.

\subsection{Late flares: delayed viscous spreading}
This is the case in which the viscous evolution of clumps begins with a significant
delay with respect to the prompt emission. In general,
there are two reasons for such a delay to occur. First, clumps may form soon after the
prompt, as in sec. \ref{sec:early}, but in outer regions of the disc where they can survive 
viscous stresses for a while before eventually migrating toward the inner region of the disc where they are
shredded. Second, disc fragmentation
may last for a while, hence clumps may be formed with some delay with respect to the initial prompt 
emission.

As time progresses after the prompt phase, the accretion rate drops, and 
hence the radius $R_{\rm inst}$ below which the Toomre parameter $Q_{\rm T}<1$ increases,
$R_{\rm inst}\propto \dot{M}^{-2/3}$. The accretion rate during the flare phase
can be estimated assuming that it tracks the mean observed luminosity 
(Lazzati et al. 2008: Margutti et al. 2011b). From a sample of 44 bursts, Margutti et al. (2011b) 
estimated a decline with an average power of $t^{-2.7}$.
Hence, during this phase,
the radius at which $Q_{\rm T}=1$ expands outward as $R_{Q_{\rm T}=1}\propto t^{1.8}$. This
means that, the later the times, the larger the disc radii at which
clumps can form. Clumps that form very far out in the disc, in regions that are very
gravitationally unstable and in which viscous torques are rather weak,
are likely to remain bound and self-gravitating for some time, migrate
inward, and only start spreading when shear and/or tidal forces become
comparable to their self-gravity.

The shear force per unit length acting on a fluid element at radius
$r$ is $f_{\nu} = \nu \Sigma r \displaystyle \frac{{\rm d} \Omega}{{\rm d}r}$. A
clump of linear size $\ell$ and mass M$_{\rm cl}$ will thus be subject
to the shear force per unit mass 
\begin{equation}
F_{\nu} = \displaystyle \frac{ \ell
  f_{\nu}}{M_{\rm cl}} = \displaystyle \frac{\ell \nu\Sigma}{M_{\rm
    cl}} ~r \left| \frac{{\rm d} \Omega}{{\rm d}r} \right| = \displaystyle
\frac{\ell \dot{M}}{2 \pi M_{\rm cl}} \Omega,
\label{eq:shear}
\end{equation} 
where we made use of
the relation $\dot{M} = 3 \pi \nu \Sigma$, valid in a steady-state
disc, and of the fact that $\Omega \propto r^{-3/2}$ for keplerian
rotation.
 
Given the self-gravity force per unit mass of the clump, $F_{\rm SG} = \displaystyle
\frac{G M_{\rm cl}}{\ell ^2}$, and the tidal force due to the central object, $F_{\rm T} =
\displaystyle \frac{G M_{*}}{r ^2} \left(\frac{\ell}{r}\right)$, 
viscous spreading of the
clump will start roughly when the ratio $V = \displaystyle
\frac{F_{\nu}+F_{\rm T}}{F_{\rm SG}} >1 $. This condition can be re-cast
as\footnote{Detailed calculations show that this criterion is correct,
  yet excessively constraining. For example, tidal disruption occurs at
  an orbital separation that is slightly larger (typically, a factor
  $\sim$ 2) than given by the condition $F_{\rm T}/F_{\rm SG} >1$. This is
  because a tidally distorted star is less self-bound, at a given
  orbital separation, than an undisturbed one, hence more prone to
  breaking.}
\begin{equation}
\label{eq:shred}
\frac{M_{*}}{M_{\rm cl}} \left(\frac{\ell}{R_{\rm s}}\right)^3 x^{-3} \left[1 + \frac{\sqrt{2}}{\pi} \frac{R_{\rm s}}{c} \frac{\dot{M}}{M_{\rm cl}} x^{3/2} \right] >  1 \,
\end{equation}
{where we have defined $x \equiv r/R_{\rm s}$.  For a given central
  object and clump mass and size, this condition defines a boundary in
  the $r$ vs. $\dot{M}$ plane between (inner) regions where clumps
  form and start spreading right away, and (outer) regions} where
newly formed clumps remain self-bound. In the latter case, they will
have to migrate inwards before being viscously/tidally shredded.  {For
  an order of magnitude estimate, we assume M$_{*} = 10~\Msol$
  and rescale the clump size based on the local Jeans length,
  $\displaystyle \lambda_J = c_{\rm s} \sqrt{\frac{\pi}{G \rho}} \sim 2
  \times 10^8 \left(\frac{x}{100}\right)^{1/2}$ cm, where the disc
  sound speed $c_{\rm s} \approx H \Omega_{\rm K} \sim 0.5 v_{\rm K}$ is typical for a
  slim disc and we have approximated the density profile for $x> 100$
  as $\rho \sim 10^9 \displaystyle
  \left(\frac{\dot{M}}{M_{\odot}/{s}}\right)
  \left(\frac{x}{100}\right)^{-2}$ g cm$^{-3}$ (Di Matteo, Perna \&
  Narayan 2002). Therefore, we assume a typical clump size $\ell \sim
  10^8$ cm. Eq.~(\ref{eq:shred}) leads us to conclude that the disc
  radius at which viscous/tidal disruption occurs is in the range
  $\sim 200-1000$ R$_s$, depending on $\dot{M}$ and on the specific values of the relevant
  physical parameters.}

A thorough discussion of all possible behaviors resulting from Eq.~(\ref{eq:shred})
is beyond the scope of this paper. Such an analysis would require
detailed numerical simulations of the process of clump formation,
cooling and fragmentation in a rapidly evolving accretion disc, where
$\dot{M}$ can drop by 2-3 orders of magnitude in hundreds of seconds.

For the discussion {presented} in Sec.~5, what matters is that clumps may form at
some outer radius $R_0$, and then migrate inwards till a radius $R_{\rm in}$
before they begin to evolve viscously. The migration timescale can be comparable 
to the viscous time, if the local gravity is dominated by the disc. On the other hand, if the
clumps are massive enough and
dominate the local potential, they will migrate inward on a timescale
$t_{\rm migr}$ which is longer than the viscous timescale by a factor
on the order of $(M_{\rm cl}/M_{\rm disc})>1$, where $M_{\rm disc}$ is
the exterior disc mass, and details depend on the precise radial
dependence of the density profile of the unperturbed disc (Syer \&
Clarke 1995).   At later times, when fallback from the envelope of the star has terminated 
and hence the remaining disc is only depleting, the condition $M_{\rm cl}/M_{\rm disc} >1$ 
may be verified, further lengthening the travel time of the clump.

In summary, as the mean accretion rate in the disc drops, clumps form
farther out and may travel in the disc before beginning to viscously
spread.  Under these conditions, their travel time may be comparable to, or longer than
the viscous timescale at $R_0$. 
Additionally, as the disc expands and its conditions change with time,
new clumps in the outer regions of the disc can form some time after the 
prompt emission. In the following, let us call $t_{\rm off}$ the offset time between
$t=0$ (defined as the time of the GRB trigger) and the time at which
viscous spreading begins.  Again, this offset could be due to a long 
migration time, or to late-born clumps, or to a combination of these
two factors.

\begin{figure}
\includegraphics[scale=0.5]{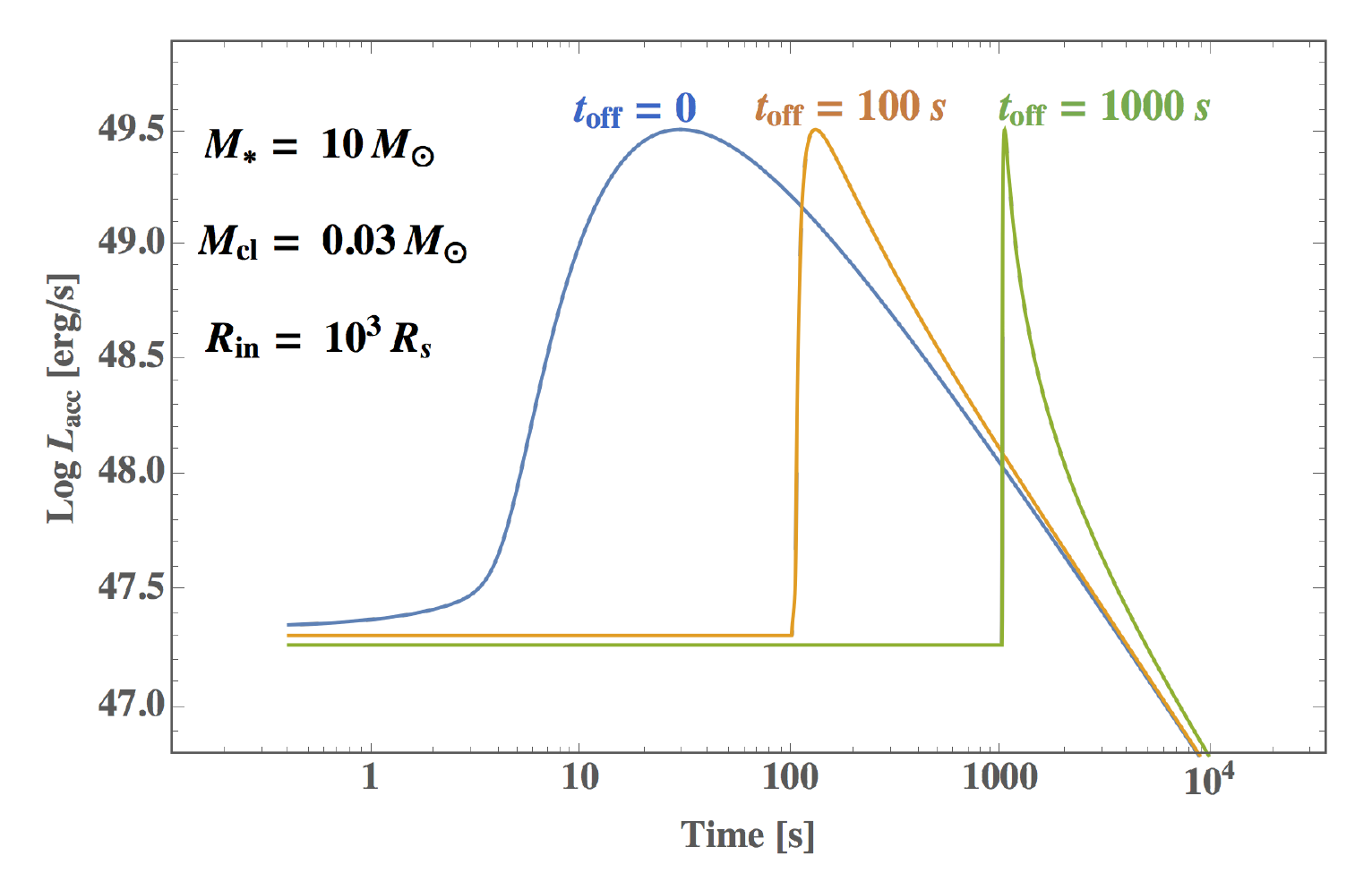}
\caption{The change in the bolometric lightcurve of a flare as its
  arrival is delayed by an offset $t_{\rm off}$, which represents the 
time between the GRB trigger and the beginning of the viscous evolution of the clump. 
The properties of the clump are the same for all three
  cases, including the radius $R_{\rm in}$ at which spreading begins;
  the only difference is in $t_{\rm off}$.}
\label{fig:case2}
\end{figure}

Figure~\ref{fig:case2} compares the observed lightcurves for a clump of identical
properties (i.e. mass, radius where spreading begins) but only
differing for the time $t_{\rm off}$ at which spreading begins.
{We note that the {\it shape} of these curves is independent of $t_{\rm off}$ --- the decay time and the rise time do not vary, hence the ratio $t_{\rm d}/t_{\rm r}$ stays constant. 
The duration $\Delta t = t_{\rm d}+t_{\rm r}$ is also unchanged; however, since the flares arrive 
at later times, the ratio $\Delta t /t_{\rm p}$ gets smaller for larger $t_{\rm off}$}. 
These properties will be further discussed and compared to the observations in Sec.~5.

\section{Comparing our model against  observations}

\subsection{lightcurve shape and $\Delta t/t_{\rm p}$ correlation}

As already mentioned in Sec.4, an important feature of our model lies
in the fact that the bolometric lightcurve produced by the viscous
spreading of the clump has a well-defined shape, which is only
marginally affected by changing the values of the physical parameters.

\begin{figure*}
\includegraphics[scale=0.4]{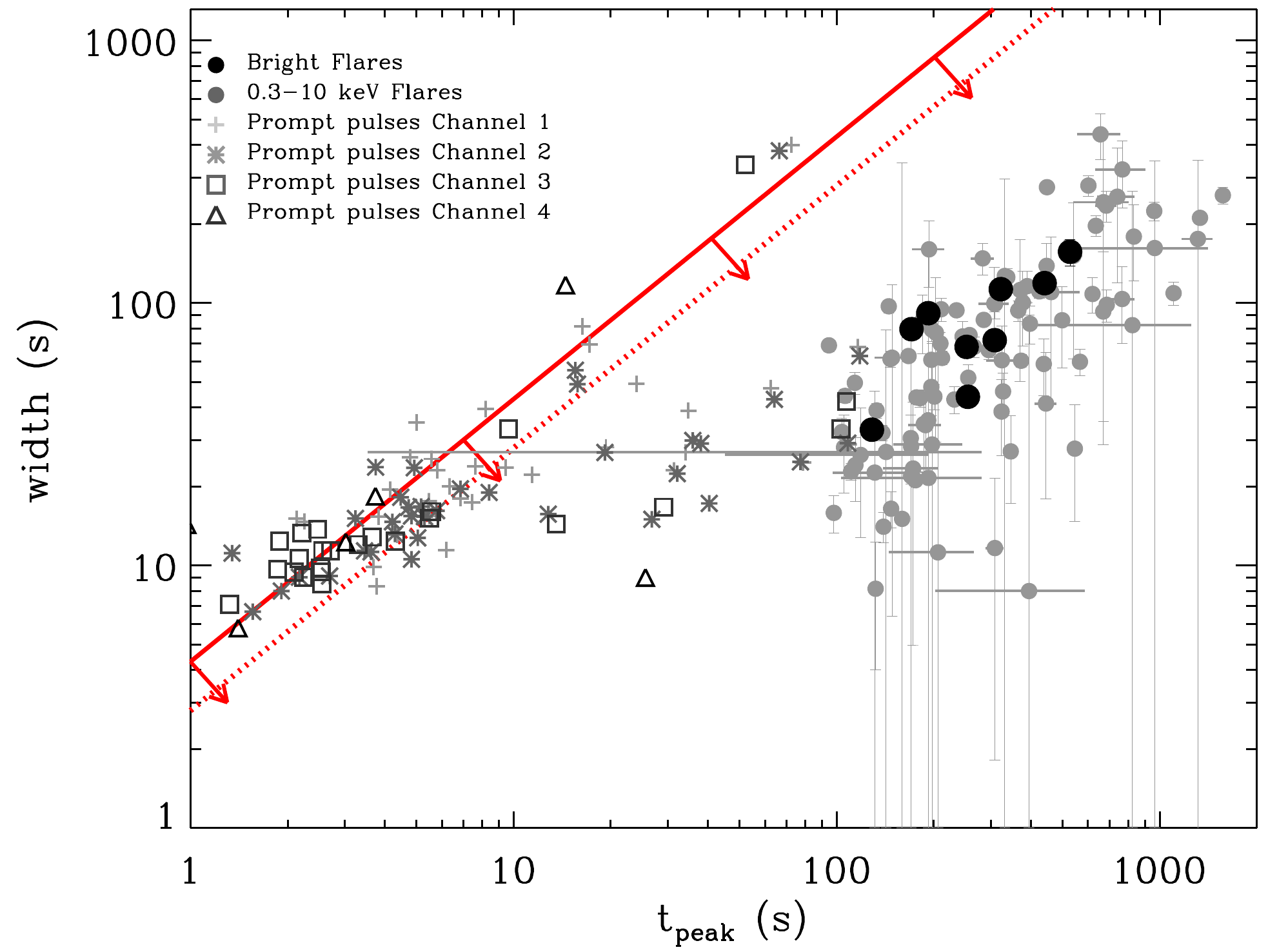}
\includegraphics[scale=0.4]{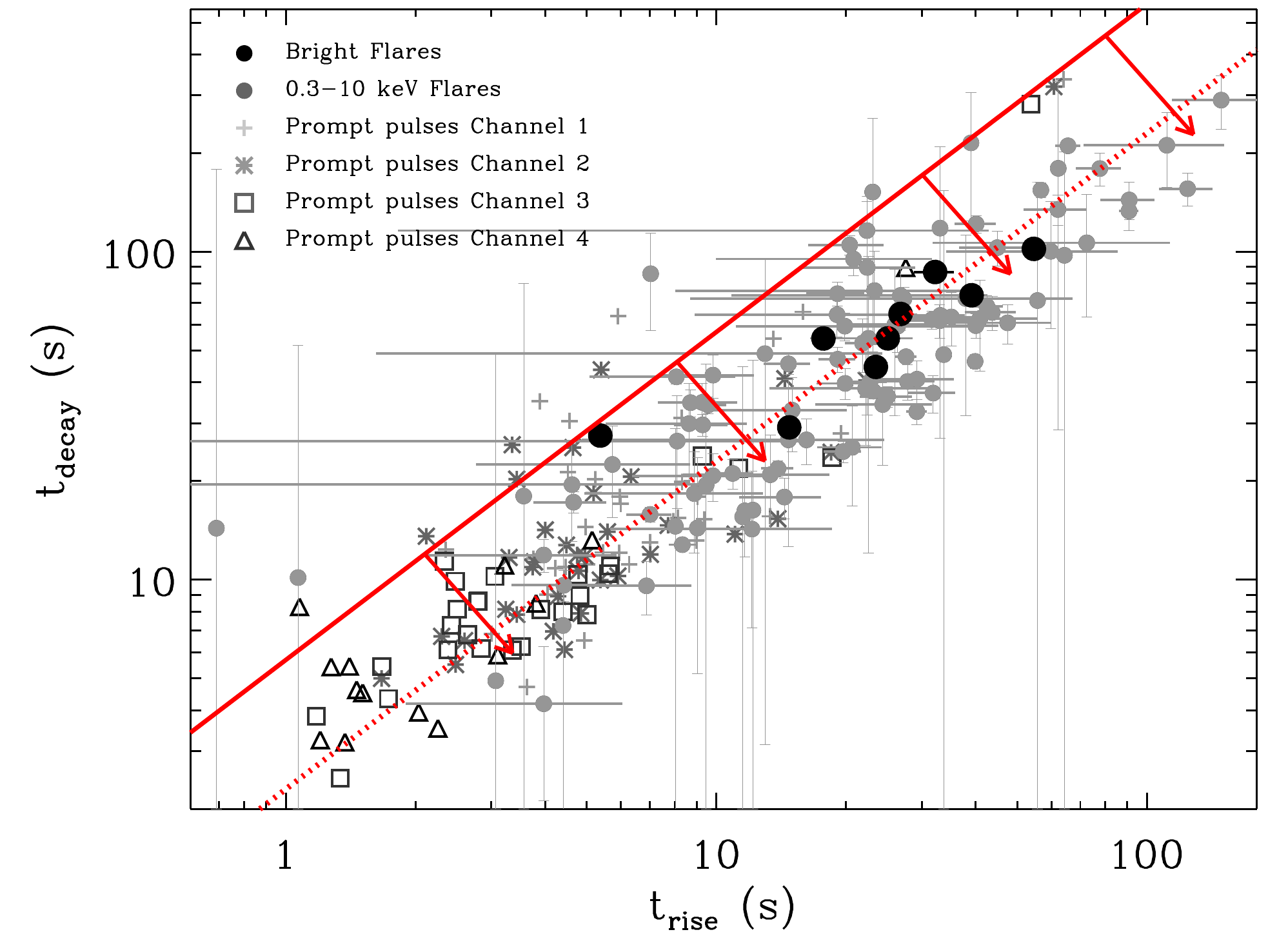}
\caption{A comparison of the flare data from the Margutti et al. (2010)
  sample with the predictions of our model. {\it Left:} Width-to-peak
  time correlation.  {\it Right:} Decay time versus rise time for each
  flare. The solid lines in both panels indicate the (highly
  constrained) model predictions from the bolometric lightcurves.
  Spectral corrections due to the finite band width introduce scatter
  and {\it decrease} the ratios ${\Delta t}/t_{\rm p}$ and $t_{\rm d}/t_{\rm r}$
  (as indicated by the arrows), albeit the latter is much more
  sensitive to spectral corrections than the former. As an example,
  the dashed lines indicate the spectral correction obtained with the
  measured parameters of a specific flare, that of GRB~060904B (see
  Sec.5.2). }
\label{fig:data}
\end{figure*}

We have studied how the values of $t_{\rm p}$, $t_1$ and $t_2$ change {as 
$\alpha$, $R_{\rm in}$, $M_{*}$ and
$R_{\rm ISCO}$ are varied} within physically plausible ranges.
Our main
conclusion is that, while {each individual timescale changes} according to the
numerical value of each of the physical parameters
(cfr. Fig.~\ref{fig:case1}), their ratios remain remarkably constant. In particular, 
the ``asymmetry parameter'' $t_{\rm d} / t_{\rm r} \approx 5.7$ and the ``width parameter'' $\Delta t/ t_{\rm p} \approx
4.3$ are insensitive to the value of $R_{\rm in}$, $\alpha$, $M_*$
and $R_{\rm ISCO}$. 
Only increasing the powerlaw index $n$ of the viscosity law from 0.5 
(an advection-dominated disc with $H/R\sim 1$ ) to 1.5 (an isothermal disc) can appreciably 
affect the lightcurve shape, producing narrower peaks with somewhat 
smaller asymmetry. However, as discussed above, GRB discs
are expected to be in the former regime.

In order to test our model predictions against actual flare data, we
consider the sample of {flares with $t_{\rm p} \lesssim~10^3$~s from
  Margutti et al. (2010),} together with the sample of long-lag, wide
prompt pulses from Norris et al. (2005).  We include this additional
sample because X-ray flares can be effectively fit with the same
functional shape of the long-lag, wide prompt pulses (Margutti et
al. 2010), allowing a direct comparison of their temporal
properties. Note that in most bursts, the prompt emission consists of
a complex forest of overlapping pulses with different temporal and
spectral properties (e.g. Norris et al. 1996). Interestingly, Norris
et al. (2005) demonstrated that prompt pulses in GRB with long
temporal lag between the high-energy and low-energy photons are
spectrally and temporally distinguished from those in bright GRBs: the
pulses are wider, with lower peak fluxes, harder low-energy spectra
and softer high-energy spectra. {We refer to these as 'prompt flares'
  hereafter.}

Fig.~\ref{fig:data} shows the flare width versus peak time (left
panel) and and the decay time versus rise time (right panel) for the
full sample described above.  It is interesting to note that the
'prompt flares' appear as a somewhat distinct population than that of
the 'standard' 
flares when it comes to $\Delta t/t_{\rm p}$: they appear to have a similar 
slope for the $\Delta t/t_{\rm p}$ correlation, but the latter flares have an 
offset, signaling a later arrival time but otherwise similar underlying physical 
properties. The similarity between the underlying phenomena is strengthened 
by the fact that no separation appears in the two populations when looking at
the $t_{\rm d}$-$t_{\rm r}$ plane: the shape of the flares, as measured by their 
{asymmetry parameter, appears to be} independent of their arrival time 
(with some scatter, which will be {discussed} in the following).

The theoretical predictions of our model can now be directly compared
with the data of Fig.~\ref{fig:data}.  As already noted, the $t_{\rm d}$
vs. $t_{\rm r}$ relation is simply given by a straight line with slope
$\approx 5.7$. Remarkably, this line appears to match the upper edge
of the scattered distribution of flares in Fig.~\ref{fig:data} (right
panel, solid line).  Additionally, our model can be represented by
another line with slope 4.3 in the $\Delta t$ vs. $t_{\rm p}$ plane,which
appears to match very well the distribution of ``early flares".
Again, the points show some scatter, but it is remarkable that our
model, with its very rigid predictions, falls exactly in the
middle of the strip of data points.

As noted above, while all data points in the $t_{\rm d}/t_{\rm r}$ plane appear to
lie along the same distribution, with a scatter corresponding to a
ratio $t_{\rm d}/t_{\rm r} \sim (1.5-6)$, the $\Delta t-t_{\rm p}$ data show a large
population of ''late flares'' that appear to be offset with respect to
our model expectations. These can be fit in the proposed picture,
however, if there is an appreciable delay (i.e.what we call $t_{\rm
  off}$, cfr. Sec.4.2) between the ''time zero'' of the observations and the time at
which they started spreading viscously in the disc. As argued in
Sec.4.2, this can indeed be the case if the clumps form at some outer
radius $R_0$, and migrate to some radius $R_{\rm in}< R_0$ before they
can begin to shred due to tidal and/or viscous torques. Or
alternatively, they simply form at some later time after the prompt
emission.

Within this scenario, the early flares ($t_{\rm p} \lesssim
10-30$ s) would thus correspond to clumps formed in the inner region
of the disc, when $\dot{M}$ is the highest and viscous spreading of
the clumps begins very close to their formation site; later clumps
will instead be formed in the outer disc, due to a strong decrease of
$\dot{M}$, where their self-gravity initially resists viscous
stresses. Such self-bound clumps would migrate inwards without
spreading, until reaching smaller radii where viscous evolution
eventually takes over: at this point, they would display the same
evolution as the early ones, just with a time delay $t_{\rm off}$. In
particular, their shape parameter $t_{\rm d}/t_{\rm r}$ would remain the same,
while the ratio between the width and arrival time, $\Delta t/t_{\rm p}$,
would be decreased by a factor $\sim t_{off}/t_{\rm r}$, giving rise to a
line parallel to that of early flares in the $\Delta t$ vs. $t_{\rm p}$
plane.

This interpretation is consistent with the fact that the first $\sim
10-20$~s of the emission roughly correspond to the typical duration of
a long GRB, during which the collapsing envelope of the star maintains
the accretion rate at a high and roughly (on average) constant
value. Once the fallback has ended, the accretion rate drops abruptly
and decays as a powerlaw (MacFadyen \& Woosley 1999; Lazzati et
al. 2008; Perna et al 2014). With a lower accretion rate, clumps form
at increasingly larger radii, and, as argued above, they need to
migrate inwards before they can be shred by the viscous forces within the
disc.

Last, it is useful to derive, within the framework of our model, a
relation for the width $\Delta t$ versus $t_p$, inclusive of the
possibility of a delayed viscous spreading $t_{\rm off}$.  Generally,
we can write $t_{\rm p} \approx t_\nu + t_{\rm off}$.  Our bolometric
light curves predict $\Delta t/t_{\rm p}\approx 4.3$ in case of prompt
spreading where $t_{\rm p} \approx t_\nu$. Therefore, if there is a
delayed spreading, we can write $\Delta t\approx 4.3\, t_{\rm p} -
4.3\, t_{\rm off}$. In the $\Delta t-t_{\rm p}$ plane these are parallel
lines of slope 4.3, and offset determined by $t_{\rm off}$.  The group
of late flares is the one for which $t_{\rm off}\gg t_\nu$; hence, since
$t_{\rm off}$ is likely to vary from burst to burst, our model
predicts a larger scatter for the ``late'' flares than for the
``early'' ones, for which $\rm t_{\rm p}\sim t_\nu$.  If, for some
bursts, $\rm t_{\rm off}\sim t_\nu$, then these would represent a subset of
flares with intermediate values of $\Delta t/t_{\rm p}$; however, on
simple statistical grounds these would be expected to be less
numerous.  The data in the left panel of Fig.~\ref{fig:data} supports
both these predictions: the late flares display significant more scatter than
the early ones, and there is a paucity of data in the intermediate
region.\footnote{Note also that, due to the several tens of seconds
  for repointing a satellite in X-rays ($\gsim 60$~s for {\it Swift})
  after the $\gamma$-ray emission has faded, there is an observational
  bias against observing flares with several tens of seconds
  duration.}.

\begin{figure*}
\includegraphics[scale=0.535]{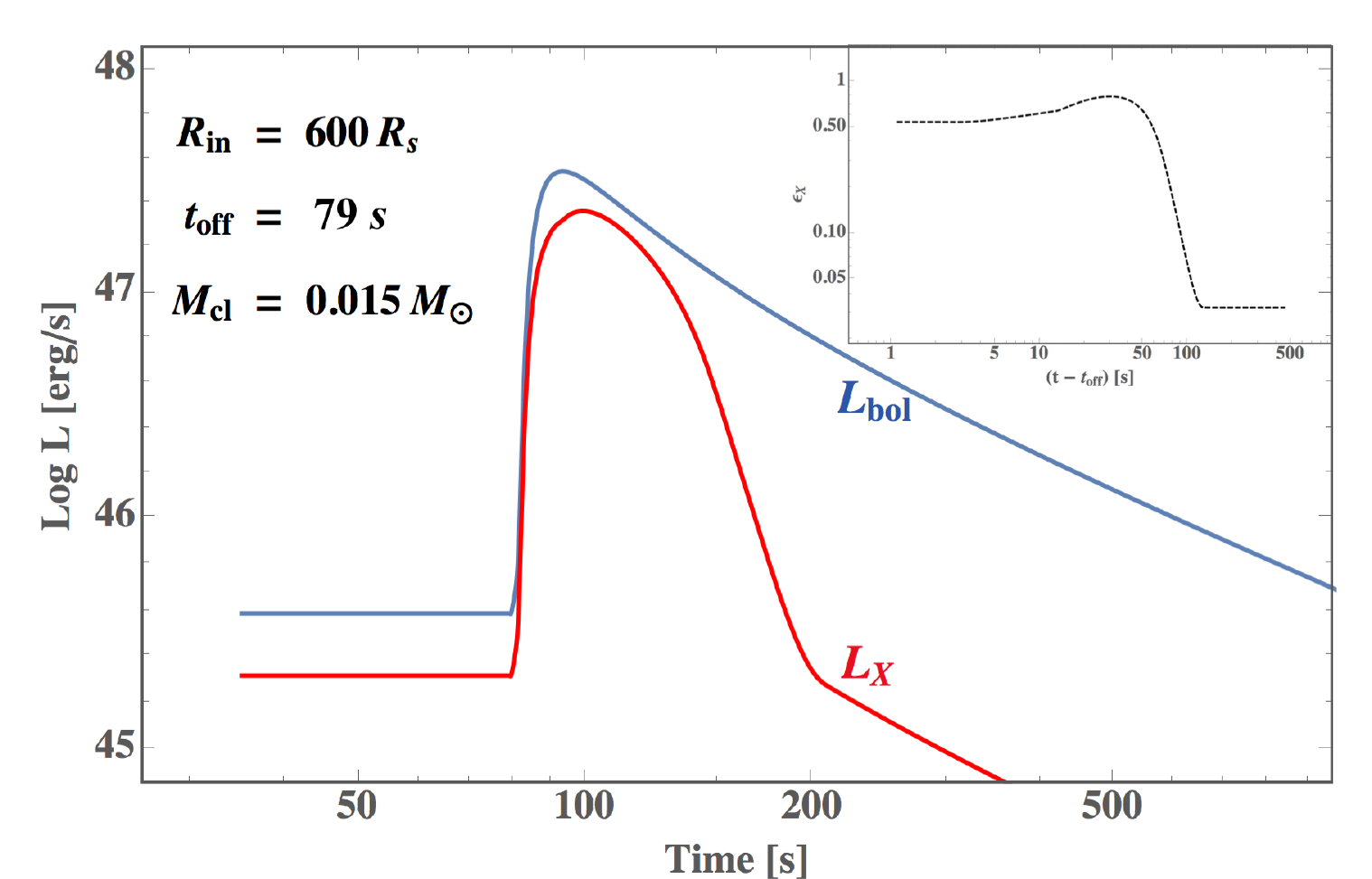}
\includegraphics[scale=0.535]{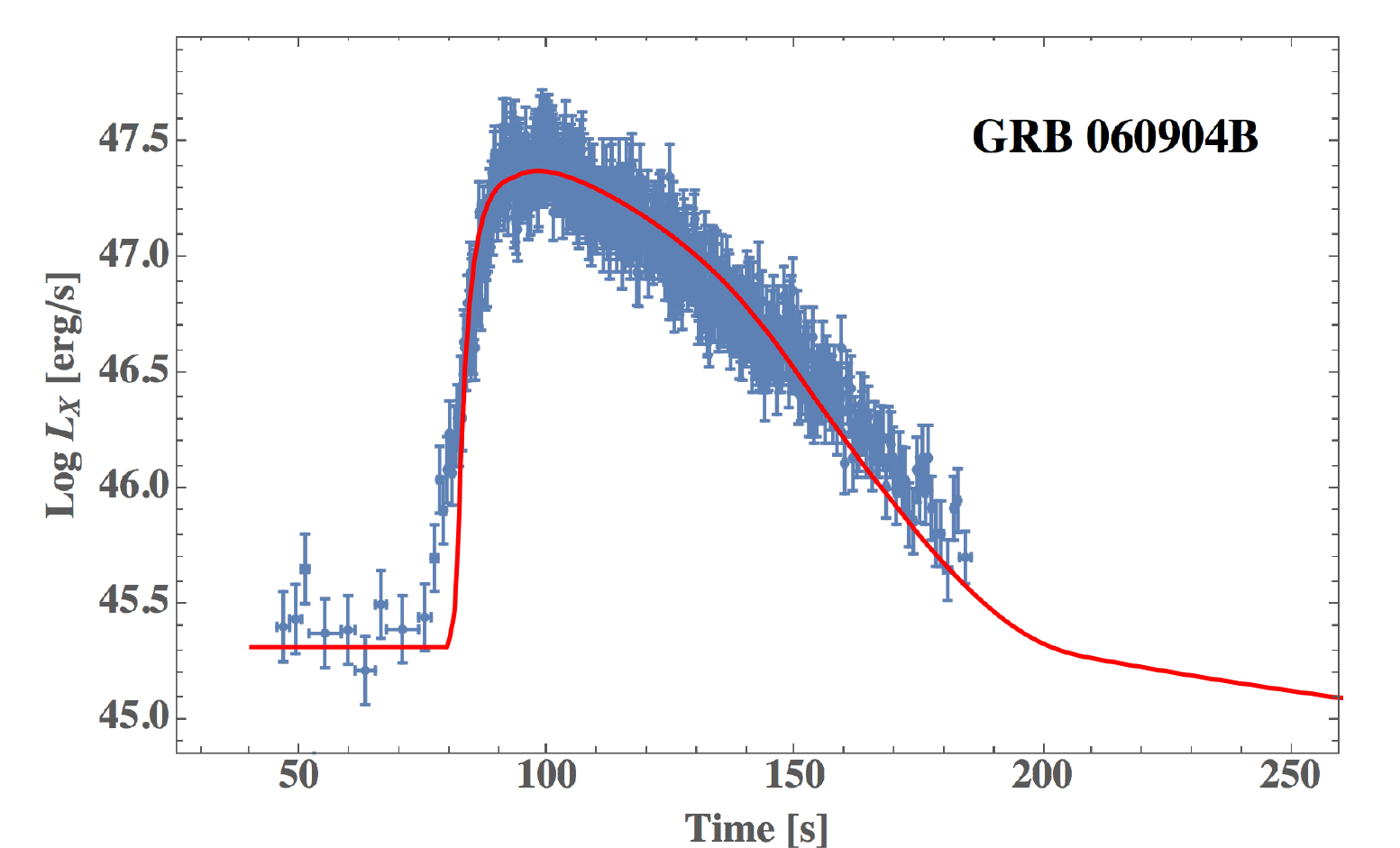}
\caption{{\it Left panel:} The bolometric lightcurve $L_{\rm bol}(t)$ (blue
  line), together with the 0.3-10~keV lightcurve $L_{\it
    X}(t)=\epsilon_{\rm X}(t)\,L_{\rm bol}(t)$ (red line) for the bright flare in
  GRB~060904B. The inset shows the time-dependent X-ray efficiency
  $\epsilon_{\it X}(t)$ derived from the {\it observed} spectrum of
  this flare. {\it Right panel:} The 0.3-10~keV luminosity of the
  flare, superimposed to the data.  Note that this is not a formal fit
  since, once the peak time of the lightcurve is
  matched with a choice of model parameters, the shape
  of the bolometric lightcurve is highly constrained, and the
  spectral correction is unique to this flare.  For these plots, we adopted $M_* = 10 M_{\odot}$, 
  a viscosity parameter $\alpha =0.1$ and power-law index $n=0.5$ for the viscosity radial profile (see Sec. \ref{sec:3}).}
\label{fig:GRB060904B}
\end{figure*}

\begin{figure*}
\includegraphics[scale=0.535]{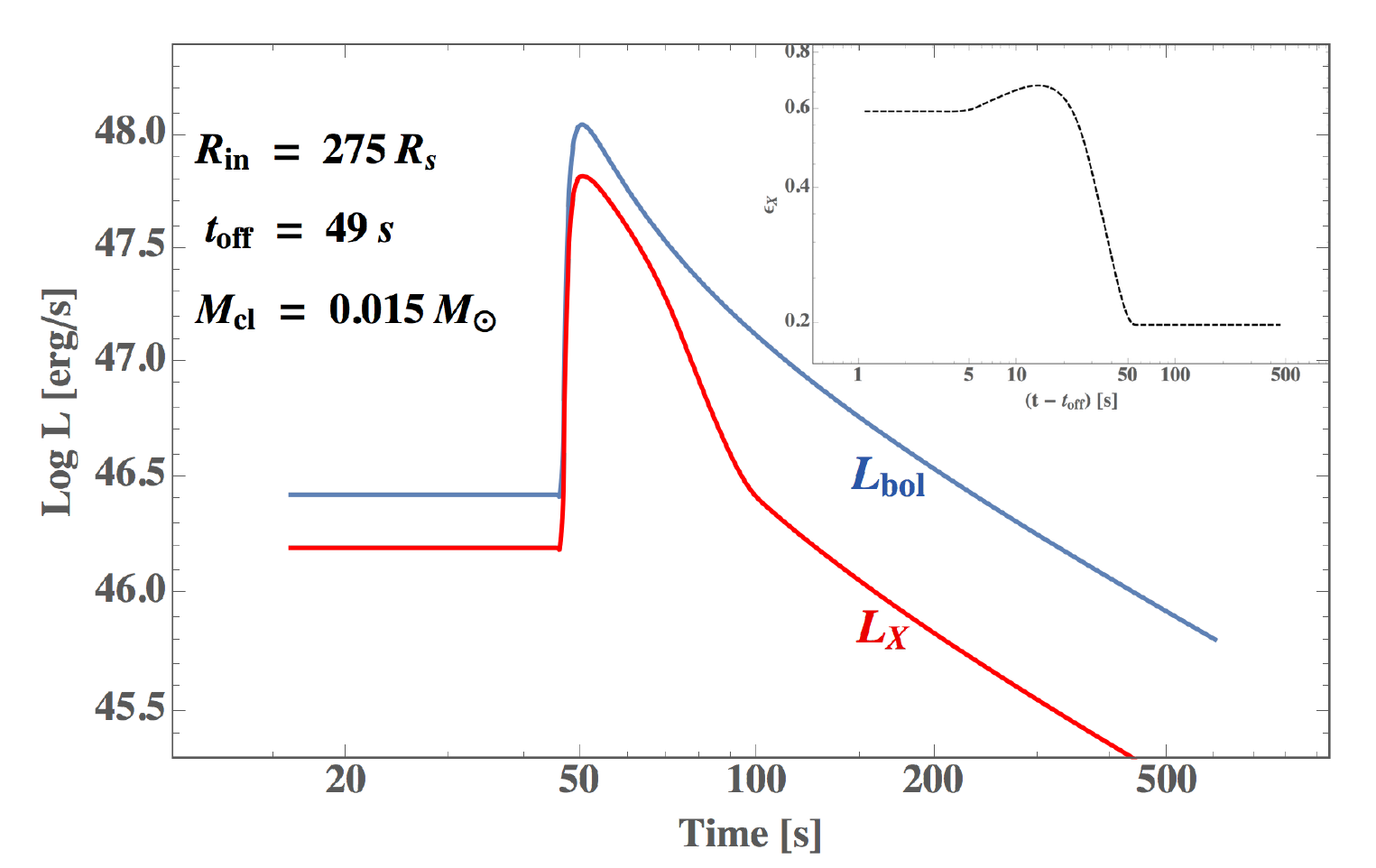}
\includegraphics[scale=0.535]{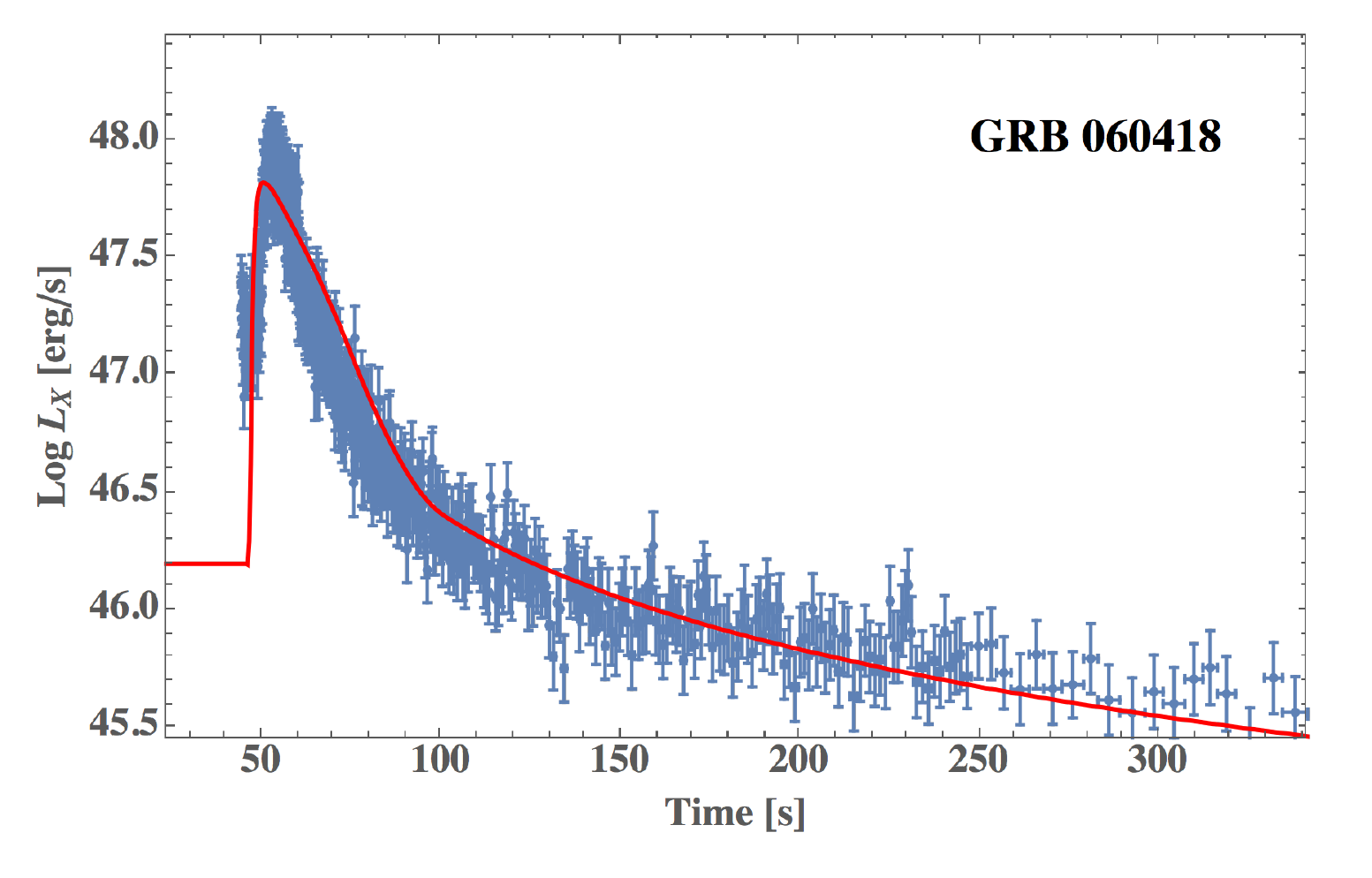}
\caption{Same as in Fig.~\ref{fig:GRB060904B}, but for the bright flare
  of GRB~060418. 
  Note that the {\it observationally derived}
  X-ray efficiency $\epsilon_{\rm X}(t)$ of this burst is lower than
  for GRB~060904B. Interestingly, a lower $\epsilon_{\rm X}(t)$ is
  needed by our model in order to reduce the highly constrained asymmetry (as measured by
  $t_{\rm d}/t_{\rm r}$) of the bolometric lightcurve to just match the observed one.}
\label{fig:GRB060418}
\end{figure*}

\subsection{Spectral Effects: Modeling of the Flares in GRB~060904B and GRB~060418}
\label{sec:spectra}

The next step in comparing our model predictions to observations calls for
the conversion from bolometric to X-ray luminosities (in the 0.3-10
keV range, where most flares have been observed),
with the goal of testing whether the model is able to: {\it i)} Explain
the scatter, in addition to the slope, in the statistical comparison
with the data of Fig.3; {\it ii)} reproduce the observed lightcurves
of individual flares.

To this end, we will rely on the observed spectra of two particularly bright flares, GRB~060904B and
GRB~060418, which were studied in detail by Margutti et al. (2010). These authors showed that flare 
spectra can be well fit by a Band function (Band et al. 1993), the spectral parameters of which change 
with time. In both flares the peak energy, $E_p$, has an exponential decay after the
flux maximum, with the same timescale of the flux decay. In addition,
Table 3 of Margutti et al. (2010) shows that the two spectral indices $\alpha$
and $\beta$ are also evolving on the same timescale, and settle at later
times at some constant values. In light of these results, for both
flares we have adopted the following analytical formulae to approximate
the evolution of their spectral parameters
\begin{equation}
E_p (t) = \left\{ \,
    \begin{IEEEeqnarraybox}[][c]{l?s}
      \IEEEstrut
      E_{p,i} & if $t \le t_{\rm max}$, \\
       E_{p,i} ~e^{-\frac{\left(t-t_{\rm max}\right)}{\tau}}  & if $t > t_{\rm max}$,  
      \IEEEstrut
    \end{IEEEeqnarraybox}
\right.
  \label{eq:epeak}
\end{equation}

and 
\begin{equation}
  \left\{
    \begin{IEEEeqnarraybox}[
      \IEEEeqnarraystrutmode
      \IEEEeqnarraystrutsizeadd{2pt}
      {2pt}
      ][c]{rCl}
      \alpha & = & \alpha_{\rm late} + a_1 e^{-\frac{(t-t_{\rm max})}{\tau}}
      \\
     \beta & = & \beta_{\rm late}+b_1  e^{-\frac{(t-t_{\rm max})}{\tau}} \, ,
    \end{IEEEeqnarraybox}
  \right.
\end{equation}
where the numerical coefficients in the two cases are 
\begin{equation}
  \begin{IEEEeqnarraybox}[
     \IEEEeqnarraystrutmode
    ]{c?c'c;v;c?c?c' c}
 & E_p {\rm [keV]} &  \tau {\rm [s]} & & \alpha_{\rm late} & \beta_{\rm late} & a_1 & b_1
    \\\hline
    060904B &  7.2 & 29 && -1 & -3 & +0.5 & +0.5\\
    060418~ &  5.7 & 20 && -1 & -2.5 & +0.1 & 0 
  \end{IEEEeqnarraybox}
\end{equation}
By means of these formulae we can calculate, as a function of time,
the fraction of the total energy that was emitted in the 0.3-10 keV,
$\epsilon_{\rm X}(t) = \left(\int_{0.3}^{10} E~N(E)~{\rm d}E \right) /
\int_{0.01}^{150} E~N(E)~{\rm d}E$. With this coefficient at hand, we can 
transform our theoretical bolometric lightcurves into X-ray
lightcurves, $L_{\rm X}(t) = \epsilon_{\rm X}(t) f_{\rm rad} L_{\rm acc}(t)$. 
The total efficiency of conversion of accreted mass into radiation is estimated to 
be typically of the order of a few percent (Giacomazzo et al. 2013 and references 
therein): as a reference value, we have adopted  $f_{\rm rad} \sim  0.01$. 

We stress how, by using an {\it observationally}-determined spectral efficiency,
our approach does not rely on any specific assumption about the (rather uncertain) emission
mechanism. However, we note that an efficiency that drops with time, as inferred by our modelling 
(see Figs. \ref{fig:GRB060904B} and \ref{fig:GRB060418}), is consistent with  the ``curvature effect" 
if the emitting region has an accelerated bulk relativistic motion (Uhm \& Zhang 2015, 2016).

For completeness, some clarification about the value of $t_{\rm max}$,
the time of the lightcurve maximum, is in order here.  Both events
considered here belong to the group of ``late flares", with $t_{\rm p} >
100$~s, most likely reflecting the enhanced migration time prior to
viscous spreading as discussed in Sec.~5.1. Hence, we calculated the
spectral efficiency, $\epsilon_{\rm X}(t)$, for bolometric lightcurves that
have a rise time matching the observed $t_{\rm r}$ of the two flares under
study (since $t_{\rm r}$, unlike $t_{\rm p}$, is unaffected by the absolute value
of the arrival time as measured from the trigger). The value of
$t_{\rm max}$ was then set to correspond to the maximum of the
bolometric lightcurves and, from these, the (0.3-10) keV lightcurve
were calculated in the way described above. Finally, we introduced in
all the functions a time offset as an arbitrary parameter, that simply
shifted the calculated X-ray lightcurves until they matched the peak
of the observed ones (while the shape was fixed).
 
To convert the (0.3 - 10) keV fluxes to the corresponding luminosities, we 
used the measured redshifts for the two flares (Margutti et al. 2010) and the 
``concordance" values for the cosmological parameters, $\Omega_M = 0.27, 
\Omega_{\Lambda} = 0.73$ and $H_0 = 70$ km s$^{-1}$ Mpc$^{-1}$. As for 
the (unknown) beaming factor of the received radiation, $f_b$, we adopted 
a typical value $\sim 0.01$ (e.g., Frail 2001).

Our results are shown in Fig.~\ref{fig:GRB060418} and \ref{fig:GRB060904B}:
for each flare, the left panel shows the bolometric lightcurve (blue),
calculated for chosen values of the physical parameters, and the
corresponding X-ray lightcurve (red) derived by adopting the observed
spectral parameters (see above). In the inset, the corresponding
spectral efficiency as a function of time is also shown. The right
panel of each figure shows the superposition of the calculated X-ray
lightcurve (red) to the data. 

We stress that the curves do not
represent formal fits: they were obtained by choosing specific values
of the physical model parameters and adopting the {\it observed}
spectra of the two flares. Nonetheless,
we emphasize an important
result: since the shape of the bolometric lightcurve is highly
constrained (as discussed in Sec.4 and 5.1) and depends very little on
the model parameters, the actual, {\it observable} shape of the light
curve becomes very sensitive to the actual spectrum of the flare. Here
we have selected two flares: one with a high spectral correction
(GRB~060409B), and another with a marginal spectral correction
(GRB~060418). The data are reproduced remarkably well by the model
{\it simpy by modeling each flare with its own   
    observationally-derived spectral efficiency.}
We found that adopting an `average spectral correction'
failed to produce satisfactory fits of individual lightcurves.
We have found this result to be an
especially attractive feature of our model.

Note that, for fixed $M_*$ and $\alpha$, the value of the initial radius $R_{\rm in}$ is
essentially determined by the rise time of the flare, leaving the mass of the clump as the only 
parameter sensitive to $f_b$ and $f_{\rm rad}$. Our results for M$_{\rm cl}$ were obtained for 
$\displaystyle \frac{f_b} {f_{\rm rad}} = 1$, and scale roughly linearly with $\displaystyle \frac{f_b} {f_{\rm rad}} $.

Finally, our exploration of the spectral effects on the bolometric light
curves has shown that, as a general feature, spectral corrections via
a time-dependent Band function tend to make the lightcurves narrower.
However, the correction to $\Delta t/t_{\rm p}$ is smaller than that to
$t_{\rm d}/t_{\rm r}$, since the time-dependent efficiencty $\epsilon_{\rm X}(t)$ drops
with time and hence it influences especially $t_{\rm d}$ (see insets in the
left panels of Figs.~\ref{fig:GRB060418} and \ref{fig:GRB060904B}).  From a
statistical point of view, this means that the theoretical, bolometric
predictions for $\Delta t/t_{\rm p}$ and $t_{\rm d}/t_{\rm r}$ (solid red lines in
Fig.~\ref{fig:data}) provide an {\it upper value} for those quantities
(modulo a small theoretical scatter with changes in the model
parameters). On the other hand the spectral corrections, which differ
from flare to flare, introduce a scatter in these quantities, which is
especially pronounced for $t_{\rm d}/t_{\rm r}$, and goes in the direction of
smaller values for the reasons explained above.  As an example, the
dashed lines in the two panels of Fig.~\ref{fig:data} show the values
of $\Delta t/t_{\rm p}$ and $t_{\rm d}/t_{\rm r}$ using the spectral efficiency derived
for GRB~060418B. 

In summary, the statistical properties of the flares are
naturally reproduced by our model.

\section{Summary and Conclusions}

We have built a quantitative model for the idea suggested by Perna et
al. (2006) to explain the flares seen in both the long and the short
GRBs. An hyperaccreting disc becomes gravitationally unstable and
fragments in its outer regions. With a sufficiently efficient cooling, the
fragments become self-gravitating. The matter clumps start migrating inwards
and, when thery reach a location at which viscous and/or tidal shears are able to
tear them apart, they begin spreading and accreting viscously.

By computing the time-evolution of the matter in the clumps and the associated
accretion luminosity, we have identified several robust features of this model:

\begin{itemize}

\item{}
If the clumps begin to spread close to the location where they were formed,
the ratio between their duration and arrival time is a robust prediction
of the model,  $\Delta t/t_{\rm p}\approx 4.3$, with an insignificant
sensitivity to all the model parameters within reasonable ranges. 

If the clumps start spreading viscously after some migration time within the disc,
then the ratio $\Delta t/t_{\rm p}$ decreases; however, the absolute value of the
width remains roughly constant.

\item{}
The asymmetry of the curve, as measured by the ratio between the decay
and the rise time, is remarkably independent of the model parameters, and has
a roughly constant value of $t_{\rm d}/t_{\rm r}\approx 5.7$; it is the same
whether the clump arrives after a migration time delay or not.

\item{} 
Spectral corrections, due to the fact that observations are
  performed in a finite energy band, play an important role in
  modifying the shape of the time-dependent bolometric luminosity
  predicted by our model.
In particular, the fast decline of the flare lightcurves does not directly track the 
 decline of the accretion power, but is best explained in terms of a marked spectral evolution. 
 This conclusion is consistent with recent findings on the effect of accelerated relativistic
 bulk motion in the emitting region (Uhm \& Zhang 2016).
 
\item{}
By using the observationally determined spectra of several flares, we
have computed the corresponding time-dependent spectral corrections to
the measured luminosity, and demonstrated that the spectral corrections
only have the effect of decreasing the width $\Delta t/t_{\rm p}$ and the ratio 
$t_{\rm d}/t_{\rm r}$ compared to the same quantities calculated without
the spectral correction. However, the correction on $t_{\rm d}/t_{\rm r}$
is larger than for $\Delta t/t_{\rm p}$.

\item{}
A comparison of the model predictions with the data shows a strong
agreement which is even more remarkable given the fact that model
is very constrained and quite insensitive to the model parameters.
Our predictions {\it naturally match the data,
they are not fit to the data}. 
More specifically, the predictions for $\Delta t/t_{\rm p}$ pass through the
data, and there is a relatively narrow scatter given by spectral corrections.
On the other hand, the theoretical line for $t_{\rm d}/t_{\rm r}$ provides
an upper limit to the data, and the relatively large scatter at lower values
is due to the more pronounced sensitivity to spectral corrections.

\item{}
While we predict that $t_{\rm d}/t_{\rm r}$ should be the same independently
of whether the clump started to spread right away or migrated first, on the
other hand the ratio $\Delta t/t_{\rm p}$ decreases if there has been
a migration time $t_{\rm off}$ before viscous spreading begins.
 
The data suggests two groups, one with $t_{\rm off}\approx 0$, and the other
with $t_{\rm off}\sim$~a few.
We identify the former group with clumps formed during the prompt phase, when
the conditions in the disc are roughly steady-state, while the latter with the
post-prompt phase, when the accretion rate begins to rapidly drop, and the
radius beyond which the disc can be unstable migrates outward. 
Clumps formed in the outer, lower density regions of the disc need to migrate
inwards before they can begin to be shred.

\end{itemize}

In addition to explaining the main properties of the flares from a
statistical base, we also modeled two specific flares with good data
coverage.  For each of them we corrected the bolometric lightcurve
(which, again, has a well-constrained shape) with the corresponding
time-dependent efficiency, determined for each of them from their own
observationally determined time-dependent spectra. 
Once each is corrected for their own spectral efficiency, the theoretical 
lightcurve matches the data remarkably well. 

\section*{Acknowledgements}
We thank Andrew MacFadyen and Riccardo Ciolfi for valuable discussions, and 
Bing Zhang and the anonymous referee for useful comments on the manuscript. 
RP acknowledges support from NASA-{\it Swift} under 
grant NNX15AR48G, and from the NSF under grant AST-1616157.

\end{document}